\documentclass[letterpaper]{article} 
\usepackage{aaai2026}  
\usepackage{times}  
\usepackage{helvet}  
\usepackage{courier}  
\usepackage[hyphens]{url}  
\usepackage{graphicx} 
\usepackage{natbib}  
\usepackage{caption} 
\usepackage{amsmath,amsfonts,amssymb}
\usepackage[linesnumbered,ruled,vlined]{algorithm2e}
\usepackage{booktabs}
\nocopyright

\title{LocDreamer: World Model-Based Learning for Joint Indoor Tracking and Anchor Scheduling}
\author{
    Geng Wang\textsuperscript{\rm 1},
    Zhouyou Gu\textsuperscript{\rm 2},
    Shenghong Li\textsuperscript{\rm 3},
    Peng Cheng\textsuperscript{\rm 4},
    Jihong Park\textsuperscript{\rm 2},\\
    Branka Vucetic\textsuperscript{\rm 1},
    Yonghui Li\textsuperscript{\rm 1}
}
\affiliations{

    \textsuperscript{\rm 1}The University of Sydney, Australia,\\
    \textsuperscript{\rm 2}Singapore University of Technology and Design, Singapore,\\
    \textsuperscript{\rm 3}CSIRO DATA61, Australia,\\
    \textsuperscript{\rm 4}La Trobe University, Australia\\

    \{geng.wang, branka.vucetic, yonghui.li\}@sydney.edu.au,
    \{zhouyou\_gu, jihong\_park\}@sutd.edu.sg,
    shenghong.li@data61.csiro.au,
    p.cheng@latrobe.edu.au
    
}

\usepackage{bibentry}

\begin{document}

\maketitle

\begin{abstract}
Accurate, resource-efficient localization and tracking enables numerous location-aware services in next-generation wireless networks. 
However, existing machine learning-based methods often require large labeled datasets while overlooking spectrum and energy efficiencies.
To fill this gap, we propose \textit{LocDreamer}, a {world model} (WM)-based framework for joint target tracking and scheduling of localization anchors.
LocDreamer learns a WM that captures the latent representation of the target motion and localization environment, thereby generating synthetic measurements to \textit{imagine} arbitrary anchor deployments. 
These measurements enable \textit{imagination-driven} training of both the tracking model and the reinforcement learning (RL)-based anchor scheduler that activates only the most informative anchors, which significantly reduce energy and signaling costs while preserving high tracking accuracy. 
Experiments on a real-world indoor dataset demonstrate that LocDreamer substantially improves data efficiency and generalization, outperforming conventional Bayesian filter with random scheduling by 37\% in tracking accuracy, and achieving 86\% of the accuracy of same model trained directly on real data.
\end{abstract}


\section{Introduction}\label{sec:section1}
Location awareness has become a cornerstone of next-generation wireless systems, supporting applications from smart buildings and industrial automation to integrated sensing and communications (ISAC)~\cite{Liu/ISAC/JSAC/2022,Trevlakis/Localization6G/2023/OJCOMS,Yang/Positioning/JSAC/2024}.
Wireless localization estimates a target's position by utilizing geometry-based measurements, e.g., distance, from multiple anchors at a single time instance, and tracking extends this across consecutive timesteps by considering temporal dynamics. 
Achieving accurate localization and tracking depends on the quality of radio-signal measurements, which often degrade under multipath and non-line-of-sight (NLoS) propagation in dynamic, complex indoor environments.
Extensive research has therefore focused on identifying and mitigating these effects from the received signals to improve tracking accuracy~\cite{Nkrow/NLoS/2024/ACMCS,Wang/Adaptive/2024/TSP}. 
Nevertheless, they typically assume that all anchors are simultaneously active to provide measurements, leading to increased energy consumption, spectrum usage, and computational overhead.
In practical wireless systems, activating all anchors at every timestep is infeasible due to limited measurement opportunities, bandwidth budgets and energy constraints. 
This motivates the need for \textit{joint tracking and anchor scheduling}\footnote{Also referred to as node selection/activation in the literature.}--dynamically selecting a subset of anchors to balance resource consumption and tracking accuracy~\cite{Win/OperationStrategy/2018/JPROC,Wang/Scheduling/2019/TNET,Haj/DQLEL/2022/TSP}.

To achieve accurate tracking and select informative anchors, model-based frameworks describe the tracking system through mathematical models, e.g., Bayesian filters, and employ heuristic or optimization-based anchor scheduling policies~\cite{Albaidhani/AnchorSelection/2019/Wiley,Zhao/CRLB/2019/TVT,Fan/CoopNS/2022/TWC,Oh/ToARSS/2023/TVT}. 
However, these static models are designed based on handcrafted assumptions that only offer interpretability and generalization under idealized conditions.
Consequently, tracking accuracy degrades and the scheduling strategies fail to adapt when the environment deviates from the assumed models. 
Moreover, their scheduling decisions still require all anchors to be active for measurements, e.g., measurements from anchors predicted to be uninformative will be discarded, and this post-processing scheme does not reduce the signaling overhead. 

Alternatively, machine learning-based frameworks can directly learn to infer target position and scheduling policies from data using neural networks~\cite{Haj/DQLEL/2022/TSP,Gomez/NN/2023/VTC,Gomez/ML/2023/MILCOM,Kim/DQN/2025/ICMLCN}, without relying on fixed parametric mathematical models.
Reinforcement learning (RL), in particular, has emerged as a promising approach that can flexibly learn strategies from collected interaction data without explicit modeling of complex environments~\cite{Haj/DQLEL/2022/TSP, Kim/DQN/2025/ICMLCN}. 
While these methods achieve high accuracy in dynamic and complex environments, they rely on extensive labeled training data to capture environment-specific features. 
Consequently, they exhibit poor generalization when anchor configurations or environmental conditions change and new measurements are unavailable~\cite{Kim/DQN/2025/ICMLCN}.
How to learn the tracking and scheduling models that generalize to unseen dynamic and complex environments without additional measurements remains an open challenge.

Recent advances in \textit{world models} (WMs) offer a powerful solution by learning compact latent representations of the environment as well as its dynamics, enabling an agent to imagine future trajectories and optimize strategies without real-world interactions~\cite{Ha/WorldModel/2018/NIPS, Hafner/MasteringWM/2025/Nature}.
Comparing to model-based RL that learns the dynamics in the observable states~\cite{Mohammadi/SSDRL/2017/JIOT,Li/USDRL/2020/JIOT}, WMs operate in a compact latent space, making them particularly effective in high-dimensional~\cite{Ha/WorldModel/2018/NIPS, Hafner/MasteringWM/2025/Nature} or complex-structured~\cite{Lee/SetTransformer/2019/ICML} environments.
By interacting with latent dynamics rather than real-world environments, WMs significantly improve sample efficiency and have achieved remarkable success in data-efficient control tasks where direct interaction with environments is expensive or unsafe, e.g., autonomous driving~\cite{Tu/WMDriving/2025/arXiv}, but its potential remains largely untapped in the field of wireless localization and tracking.

In this paper, we introduce LocDreamer, a WM-based learning framework for joint tracking and anchor scheduling to achieve high tracking accuracy and resource efficiency in dynamic indoor environments, while enabling effective adaptation to new environments without additional measurements.
Our contributions are summarized as follows:
\begin{itemize}
    \item \textbf{Joint tracking and scheduling based on WM}: 
    We formulate joint tracking and anchor scheduling as a unified maximum likelihood estimation problem based on WM. 
    Specifically, the problem jointly optimizes 1) a WM that learns the most likely dynamics in localization system to accurately track the target, and 2) an anchor scheduling policy that selects the most informative anchors, both using measurements generated by the WM.
    \item \textbf{Imagination-driven learning using WM}: 
    We design the learning framework, LocDreamer, to train the WM and the scheduling policy for joint tracking and anchor scheduling.
    The WM is first pre-trained using data from a well-measured source environment to learn the dynamics of the localization system. 
    It then imagines the dynamics and generates synthetic measurements for unseen anchor configurations, enabling self-supervised learning of joint tracking and scheduling without additional data collection. 
    \item \textbf{Evaluations using real-world dataset}: 
    Through extensive experiments, we demonstrate that LocDreamer generalizes well to unmeasured environments with new anchor configurations without any additional measurements. 
    The imagined WM and scheduling policy outperform model–based tracking with random scheduling strategy by 37\%, while achieving 86\% of the tracking accuracy of the same WM trained directly on real-world data.
    \end{itemize}


\section{System Model}
\subsection{Tracking and Scheduling Models}
We consider an indoor tracking scenario in which a target moves within a two-dimensional space $\mathcal{M} \subset \mathbb{R}^2$, covered by a set of anchors $\mathcal{A}=\left\{1,\,\cdots,\,A\right\},\,A\geq3$ at fixed known positions $\mathbf{p}^k = \left[x^k,\,y^k\right] \in \mathcal{M}$, $k \in \mathcal{A}$. 
At each discrete time step $t\in\left\{1,\,\cdots,\,T\right\}$, the target initiates ranging requests to a selected subset of anchors $\mathcal{K}_t \subseteq \mathcal{A}$ with $|\mathcal{K}_t|=K_t,\,3\leq K_t \leq A$, and receives corresponding distance measurements $\mathbf{d}_t =\left[d_t^k\right]_{k\in\mathcal{K}_t}\in\mathbb{R}^{K_t}$ from those $K_t$ anchors. 
The measured distance $d_t^k$ from each individual anchor is modeled as
\begin{equation}\label{eq:distance}
    d_t^k = \big\|\mathbf{p}_t-\mathbf{p}^k\big\| + n_t^k,
\end{equation}
where $\mathbf{p}_t=\left[x_t,\,y_t\right]\in\mathcal{M}$ is the target position and $n_t^k$ represents measurement noise. 

Before each ranging request, the anchor scheduling model determines which set of anchors $\mathcal{K}_t$ should be activated for measurements by outputting a binary scheduling policy vector $\boldsymbol{\alpha}_t=\left[\alpha_t^1,\,\cdots,\alpha_t^A\right]\in\left\{0,1\right\}^A$, where ${\alpha}_t^k=1$ if anchor $k$ is activated and ${\alpha}_t^k=0$ otherwise. 
The active set is therefore $\mathcal{K}_t=\left\{k\,|\,\alpha^k_t=1\right\}$ with $\sum_{k=1}^A\alpha_t^k={K}_t$.
Since distance measurements $\mathbf{d}_t$ can be noisy and erroneous due to multipath and NLoS propagation conditions~\cite{Wang/Adaptive/2024/TSP}, finding the optimal set of anchors $\mathcal{K}_t$ can improve both tracking accuracy and resource efficiency. 

\subsection{Unified Objective for Joint Tracking and Scheduling}
The goal is to maximize tracking accuracy subject to the constraint of scheduling $K_t$ anchors per timestep. 
In classical supervised-learning based tracking, this is equivalent to minimizing the position estimation error between the predicted and ground-truth positions. 
Since we do not assume access to labeled ground truth positions, we instead maximize the marginal likelihood of distance measurements as a surrogate for tracking accuracy
\begin{equation}\label{eq:marginal likelihood}
    \begin{aligned}
        \max_{\theta,\,\boldsymbol{\alpha}_{1:T}} \quad 
        & \log p_{\theta}\!\left(\mathbf{d}_{1:T}\,|\,\boldsymbol{\alpha}_{1:T}\right)
        \\
        \text{s.t.} \quad 
        & \sum_{k=1}^A\alpha_t^k={K}_t, \quad \forall t,
    \end{aligned}
\end{equation}
which measures how well $p_{\theta}$ explains the observed measurements under the scheduled anchors.
Intuitively, the anchor scheduling policy is optimized to maximize this likelihood, thereby favoring anchors that yield more informative and accurate measurements.

The marginal likelihood of the measurements $\mathbf{d}_t$ given anchor selections $\boldsymbol{\alpha}_t$ in~\eqref{eq:marginal likelihood} is linked to the tracking problem through the latent target state defined as $\mathbf{z}_t\triangleq \left[\mathbf{p}_t,\,{v}^x_t,\,{v}^y_t\right]\in\mathbb{R}^{Z=4}$, which also includes the velocities at $x$ and $y$ directions as kinematic states.
The tracking process is conventionally described as a discrete-time continuous-value state space model (SSM) characterized by the following state {transition} and {observation} functions 
\begin{subequations}
    \begin{align}
        \mathbf{z}_t &\sim f_{\theta_\mathbf{z}}\left(\mathbf{z}_{t-1}\right), 
        \quad \mathbf{z}_t \in \mathbb{R}^Z,
        \\
        \mathbf{d}_t &\sim g_{\theta_\mathbf{d}}\left(\mathbf{z}_t,\,\boldsymbol{\alpha}_t\right), 
        \quad \mathbf{d}_t \in \mathbb{R}^{K_t}. 
    \end{align}
\end{subequations}
where the target state $\mathbf{z}_t$ evolves following the first-order Markovian assumption and current distance measurement $\mathbf{d}_t$ is generated from the current latent state $\mathbf{z}_t$ and selected anchors $\boldsymbol{\alpha}_t$.
Applying the above, the marginal likelihood in~\eqref{eq:marginal likelihood} can be factorized as
\begin{equation}\label{eq:factorization}
\begin{aligned}
    \log & p_{\theta}\!\left(\mathbf{d}_{1:T}\,|\,\boldsymbol{\alpha}_{1:T}\right) 
    \\
    &= 
    \log \int p_{\theta}\!\left(\mathbf{d}_{1:T},\,\mathbf{z}_{1:T}\,|\,\boldsymbol{\alpha}_{1:T}\right) \mathrm{d}\mathbf{z}_{1:T}
    \\
    &= \log \int \prod_{t=1}^T g_{\theta_\mathbf{d}}\!\left(\mathbf{d}_t\,|\,\mathbf{z}_t,\,\boldsymbol{\alpha}_t\right)
    f_{\theta_\mathbf{z}}\!\left(\mathbf{z}_t\,|\,\mathbf{z}_{t-1}\right)  \mathrm{d}\mathbf{z}_{1:T},
\end{aligned}
\end{equation}
where the scheduling action $\boldsymbol{\alpha}_t$ affects the measurement likelihood, which in turn influences position inference.
However, conventional SSMs with simplified first-order assumptions cannot capture the complex hidden dynamics of the tracking system.
To address this issue, deep SSMs (DSSMs)~\cite{Girin/DVAE/2021/FTML,Gedon/DSSMSysId/2021/IFAC,Wang/DSSM/2025/TSP} have been developed to replace SSM functions with neural networks with an additional recurrence hidden state $\mathbf{h}_t$ for improved expressivity. 
As a result, we rewrite \eqref{eq:factorization} as\footnote{The recurrence hidden state $\mathbf{h}_t$ is a deterministic function of other latent variables and is marginalized for clarity of presentation throughout the paper.}
\begin{equation}\label{eq:factorization_dssm}
\begin{aligned}
    &\log p_{\theta}\!\left(\mathbf{d}_{1:T}\,|\,\boldsymbol{\alpha}_{1:T}\right)\\
    =&\log \int \prod_{t=1}^T g_{\theta_\mathbf{d}}\!\left(\mathbf{d}_t\,|\,\mathbf{z}_t,\,\mathbf{h}_t,\boldsymbol{\alpha}_t\right)f_{\theta_\mathbf{z}}\!\left(\mathbf{z}_t\,|\,\mathbf{z}_{t-1},\mathbf{h}_{t}\right) \mathrm{d}\mathbf{z}_{1:T}.
\end{aligned}
\end{equation}
\begin{figure}[t]
	\centering
    {\includegraphics[width=0.45\textwidth]{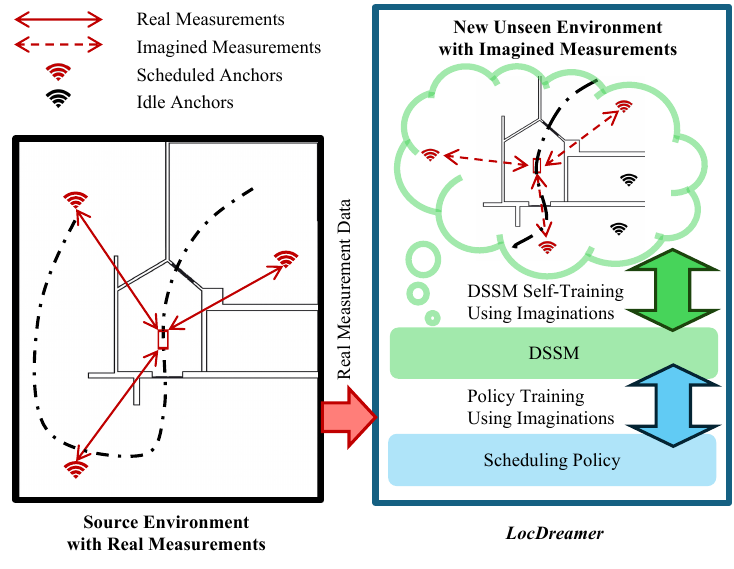}}
	\caption{Illustration of the proposed LocDreamer.}
	\label{fig:system diagram}
\end{figure}
This architecture uses the latent state $\mathbf{z}_t$ and the deterministic hidden state $\mathbf{h}_t$ to represent the system at each timestep $t$. 
Rolling out the generative model yields imagined trajectories, analogous to the imagination process observed in WM-based learning.
From imagination, we reconstruct measurements for arbitrary anchor deployments, enabling training of both the DSSM tracker and the scheduling policy in unseen environments without additional real data. An overview of the proposed imagination-based training framework, LocDreamer, is shown in Fig.~\ref{fig:system diagram}. We detail the DSSM and the imagination-driven training in the next section.

\begin{figure}[!t]
	\centering
    {\includegraphics[width=0.35\textwidth]{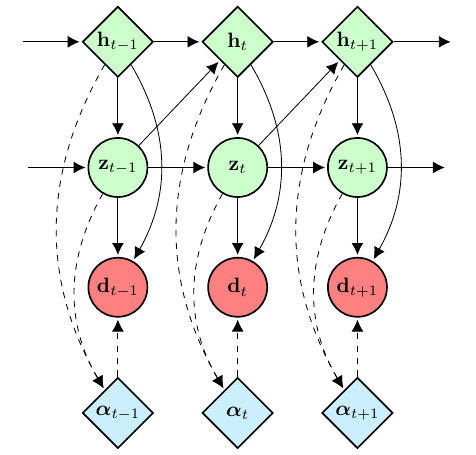}}
	\caption{Graphical model of LocDreamer. 
    Diamond and circle represent deterministic and stochastic variables. Arrows indicate conditional dependencies. Solid and dashed lines represent the DSSM modeling the tracking system and the scheduling decision making. Green, red and blue highlight DSSM, observation and anchor scheduler components, respectively.}
	\label{fig:graphical model}
\end{figure}
\section{Proposed LocDreamer}
To address the joint tracking and anchor scheduling problem formulated in~\eqref{eq:marginal likelihood}, we propose LocDreamer, a WM-based learning framework that integrates DSSM for probabilistic tracking and imagination-driven RL scheduler training within a single framework.
The DSSM learns the latent/hidden dynamics of the target and the environment, while the RL agent learns optimal scheduling policies.
This imagination capability allows the DSSM to be first pre-trained in a source environment, and then used to generate synthetic measurements for unseen anchor configurations in new environments. 
Using the generated measurements, LocDreamer self-supervises the training of both the DSSM and the RL scheduler, without requiring additional real-world measurements in the new environments.
Since the overall learning objective is to maximize the marginal likelihood in~\eqref{eq:marginal likelihood}, DSSM approximates the likelihood of the measurements from scheduled anchors, and the RL scheduler learns to select anchors that provide the highest-likelihood measurement in DSSM, thus improving tracking accuracy while reducing resource consumption.
The graphical model of LocDreamer is shown in Fig.~\ref{fig:graphical model}. 

\begin{figure*}[!ht]
\centering
\normalsize
\begin{equation}\label{eq:elbo}
    \mathcal{L}^{\text{ELBO}}\!
    = \!\sum_{t=1}^T
    \Bigg\{\!\!
    \underbrace{\mathbb{E}_{q_{\phi}}\!
    \Biggl[
    \sum_{k\in\mathcal{K}_t} \! \log g_{\theta_{d}}\!\!\left({d}^k_{t}\,|\,\mathbf{z}^\text{posterior}_t,\,\mathbf{h}_t,\,\mathbf{p}^k\right)
    \Biggr]}_{L^\text{recon}_{t}}
    \!-\! 
    \underbrace{\text{KL}
    \biggl(\! q_{\phi}\!\! \left(\mathbf{z}^\text{posterior}_{t}\,|\,\mathbf{z}_{t-1}^\text{posterior},\,\mathbf{h}_{t},\,\mathbf{o}\!\left(\boldsymbol{\alpha}_t\right)\right) \Big\|\,f_{\theta_\mathbf{z}}\!\! \left(\mathbf{z}^\text{prior}_t\,|\,\mathbf{z}_{t-1}^\text{posterior},\,\mathbf{h}_{t}\right)
    \biggr)}_{L^\text{dyn}_{t}} 
    \!\Bigg\}.
\end{equation}
\noindent\rule{\textwidth}{0.5pt}\\[0ex]
\end{figure*}
\subsection{Deep State Space Model in LocDreamer}
The DSSM backbone~\cite{Girin/DVAE/2021/FTML,Gedon/DSSMSysId/2021/IFAC,Wang/DSSM/2025/TSP} of our LocDreamer integrates stochastic recurrent latent dynamics, physics-guided modeling and variational inference to learn the tracking dynamics.
It consists of the following components: a sequence model to capture temporal dependencies of the hidden states; a dynamic model to predict the prior latent state in the next step; an encoder model to infer the posterior latent state from measurements; and a decoder model to reconstruct measurements from the latent state. 
Their structures are detailed as follows.

\subsubsection{Sequence Model}
We use recurrent neural network (RNN) as the sequence model that acts as the memory of DSSM. 
The RNN deterministic hidden state $\mathbf{h}_t\in\mathbb{R}^H$ evolves as
\begin{equation}\label{eq:rnn sequence}
    \mathbf{h}_t = r_{\theta_\mathbf{h}}\left(\mathbf{z}_{t-1}^\text{posterior},\,\mathbf{h}_{t-1}\right),
\end{equation}
where $r_{\theta_\mathbf{h}}\left(\cdot\right)$ is a recurrent function parametrized by $\theta_\mathbf{h}$, and $\mathbf{h}_t$ depends on both the past posterior estimation of the latent state $\mathbf{z}_{t-1}$ and the past RNN hidden state $\mathbf{h}_{t-1}$.

\subsubsection{Dynamic Model}
We model the target's prior latent state $\mathbf{z}_t^\text{prior}$ as a Gaussian-distributed random variable
\begin{equation}
    \mathbf{z}_t^\text{prior}
    \sim\mathcal{N}\left(\boldsymbol{\mu}_{\mathbf{z},t}^{\text{prior}},\,\text{diag}\left\{\left(\boldsymbol{\sigma}_{\mathbf{z},t}^{\text{prior}}\right)^2\right\} \right), 
\end{equation}
with mean vector $\boldsymbol{\mu}_{\mathbf{z},t}^{\text{prior}}$ and a diagonal covariance matrix $\text{diag}\left\{\left(\boldsymbol{\sigma}_{\mathbf{z},t}^{\text{prior}}\right)^2\right\}$. 
These parameters are generated by the target dynamic model that consists of a physics-based approximation and a learning-based correction as
\begin{equation}\label{eq:target dynamic}
\begin{aligned}
    \left[\boldsymbol{\mu}_{\mathbf{z},t}^{\text{prior}},\,\boldsymbol{\sigma}_{\mathbf{z},t}^{\text{prior}}\right] 
    &= f_{\theta_\mathbf{z}}\left(\mathbf{z}_{t-1}^\text{posterior},\,\mathbf{h}_t\right)
    \\
    &=f^\text{phy}\left(\mathbf{z}_{t-1}^\text{posterior}\right)+\text{MLP}_{\theta_\mathbf{z}}\left(\mathbf{z}_{t-1}^\text{posterior},\,\mathbf{h}_{t}\right),
    \end{aligned}
\end{equation}
where $f^\text{phy}\left(\cdot\right)$ is an arbitrary motion model~\cite{Li/DynamicModelSurvey/2003/TAES} and $\text{MLP}_{\theta_\mathbf{z}}\left(\cdot\right)$ is a multi-layer perceptron (MLP) parametrized by $\theta_\mathbf{z}$ that learns a data-driven residual compensating model discrepancies.

\subsubsection{Encoder Model}
At each timestep $t$, when encoding the measurements from either real or imagined environments, the model receives pairs of distance measurement ${d}_t^k$ and its corresponding anchor position $\mathbf{p}^k$ as observations $\mathbf{o}\!\left(\boldsymbol{\alpha}_t\right)\triangleq\left\{\left(d^k_t,\,\mathbf{p}^k\right)\,|\,\alpha^k_t=1, k=1,\dots,A \right\}$ from scheduling action $\boldsymbol{\alpha}_t$.
Similarly, we model the target posterior state $\mathbf{z}_t^\text{posterior}$ as a Gaussian-distributed random variable
\begin{equation}
    \mathbf{z}^\text{posterior}_{t}
    \sim \mathcal{N}\left(\boldsymbol{\mu}_{\mathbf{z},t}^{\text{posterior}},\,\text{diag}\left\{\left(\boldsymbol{\sigma}_{\mathbf{z},t}^{\text{posterior}}\right)^2\right\}\right). 
\end{equation}
And we apply variational Bayesian inference~\cite{Kingma/VAE/2022/arXiv} to approximate the true posterior as 
\begin{equation}\label{eq:encoder}
    \left[\boldsymbol{\mu}_{\mathbf{z},t}^{\text{posterior}},\,\boldsymbol{\sigma}_{\mathbf{z},t}^{\text{posterior}}\right] 
    = q_\phi\left(\mathbf{z}_{t-1}^\text{posterior},\,\mathbf{h}_{t},\,\mathbf{o}\!\left(\boldsymbol{\alpha}_t\right)\right), 
\end{equation}
where $q_\phi\left(\cdot\right)$ is the variational inference function parameterized by $\phi$.
We implement $q_\phi\left(\cdot\right)$ following the set transformer encoder approach~\cite{Lee/SetTransformer/2019/ICML} to ensure permutation invariance with respect to the number and order of active anchors in the observations $\mathbf{o}\!\left(\boldsymbol{\alpha}_t\right)$.

\subsubsection{Decoder Model}
The decoder model reconstructs distance measurements by embedding physics-based distance model in~\eqref{eq:distance} with a learnable noise model.
The reconstructed distance measurement $\hat{d}^{k}_t$ of the $k$-th anchor is modeled as a Gaussian-distributed random variable as
\begin{equation}\label{eq:decoder_dis}
    \hat{d}^{k}_t
    \sim \mathcal{N}\left(\mu_{d,t}^{k},\,\left(\sigma_{d,t}^{k}\right)^2\right), \quad \forall k \in \mathcal{K}_t,
\end{equation}
whose parameters are inferred based on the posterior estimation of the latent state as
\begin{subequations}\label{eq:decoder}
\begin{align}
    \mu_{d,t}^{k}
    &= \|\tilde{\mathbf{p}}^{\text{posterior}}_t - \mathbf{p}^k\|,\quad \forall k \in \mathcal{K}_t,\label{eq:decoder:a}\\
    \sigma_{d,t}^{k}
    &=
    \text{MLP}_{\theta_{d}}\left(\tilde{\mathbf{z}}^\text{posterior}_t,\,\mathbf{h}_t,\,\mathbf{p}^k\right), \quad\forall k \in \mathcal{K}_t. \label{eq:decoder:b}
\end{align}
\end{subequations}
Here, \eqref{eq:decoder:a} uses physics-based distance model in~\eqref{eq:distance} to compute the mean of the reconstructed distance and~\eqref{eq:decoder:b} uses a MLP parameterized by $\theta_d$ to learn the measurement noise with reparameterized sample $\tilde{\mathbf{z}}^\text{posterior}_t$ from ${\mathbf{z}}^\text{posterior}_t$.
We write the above reconstruction process in~\eqref{eq:decoder_dis},~\eqref{eq:decoder:a}~and~\eqref{eq:decoder:b} as $\hat{d}^k_t \sim g_{\theta_{d}}\left(\mathbf{z}^\text{posterior}_t,\,\mathbf{h}_t,\,\mathbf{p}^k\right)$.

\subsubsection{Loss of DSSM}
Note that direct maximization of~\eqref{eq:marginal likelihood} is intractable because it requires the integration over all possible latent state sequences $\mathbf{z}_{1:T}$ in~\eqref{eq:factorization_dssm}. 
Therefore, to train the DSSM, we maximize the variational \textit{evidence lower bound} (ELBO) of~\eqref{eq:marginal likelihood} in~\eqref{eq:elbo} as an alternative objective that can be computed based on the DSSM components in~\eqref{eq:rnn sequence},~\eqref{eq:target dynamic},~\eqref{eq:encoder} and~\eqref{eq:decoder}~\cite{Girin/DVAE/2021/FTML, Wang/DSSM/2025/TSP}.
Here, the reconstruction loss $L^\text{recon}_t$ computes the conditional log-likelihood of the measurements, encouraging the model to reproduce realistic distance distributions, while the dynamic loss $L^\text{dyn}_t$ regularizes posterior approximation from prior dynamics, enabling coherent imagination.

\subsection{RL-Based Anchor Scheduling in LocDreamer}
With the learned DSSM that models the tracking system dynamics, we can now optimize the anchor scheduling policy to select informative anchors that maximize the measurement likelihood.
We use an actor-critic (AC) model to train our scheduling policy entirely based on the imagined measurements generated by the DSSM. 
\subsubsection{Actor Model}
The actor is the stochastic scheduling policy that makes the actor selections $\boldsymbol{\alpha}_t$ based on the current latent and hidden state of the DSSM
\begin{equation}\label{eq:action}
    \boldsymbol{\alpha}_t \sim \pi_{\phi_{\boldsymbol{\alpha}}}\!\left( \mathbf{s}_t \right), \quad \forall t,
\end{equation}
where $\mathbf{s}_t = \left(\mathbf{z}^\text{prior}_t,\,\mathbf{h}_t\right)$ is the state input of the actor at timestep $t$ and $\phi_{\boldsymbol{\alpha}}$ is the parameter of the actor model.

\subsubsection{Critic Model}
The critic learns to evaluate the expected reward to guide the actor based on the rewards. 
The reward $R_t=L_t^\text{recon}$ in~\eqref{eq:elbo} is designed as the reconstruction loss as the policy is rewarded for selecting anchors that yield high measurement likelihood, i.e., anchors that preserve tracking accuracy. 
The critic predicts the value over the future discounted rewards as
\begin{equation}
    V_{\phi_{V}}\!\left(\mathbf{s}_t\right)=\mathbb{E}\left[\sum_{\tau=t}^T\gamma^{\tau-t}R_\tau\right],
\end{equation}
where $\phi_{V}$ is the parameter of the value function, and $\gamma\in\left(0,1\right)$ is the discount factor. 

\subsubsection{Loss of Actor Critic}
The actor learns to maximize the rewards with each anchor scheduling strategy
\begin{equation}
    L_t^\text{actor} =  - \mathbb{E}_{\sim \pi_{\phi_{\boldsymbol{\alpha}}},\text{DSSM}}[\log \pi_{\phi_{\boldsymbol{\alpha}}}\left({\boldsymbol{\alpha}}_t\,|\,\mathbf{s}_t\right) Adv\left(\mathbf{s}_t,\,{\boldsymbol{\alpha}}_t\right)],
\end{equation}
where minimizing $L_t^\text{actor}$ is equivalent to maximizing the expected return under the policy parameterized by $\phi_{\boldsymbol{\alpha}}$ when interacting with the learned DSSM,
and $Adv\left(\mathbf{s}_t,\,\boldsymbol{\alpha}_t\right) = \sum_{\tau=t}^T\gamma^{\tau-t}R_\tau-V_{\phi_{V}}\left(\mathbf{s}_t\right)$ is the advantage function representing the advantage of taking action $\boldsymbol{\alpha}_t$ in state $\mathbf{s}_t$.
The critic learns to evaluate the value by minimizing the error between the predicted value and observed return as
\begin{equation}
    \begin{aligned}
    L_t^\text{critic} &= \mathbb{E}_{\sim \pi_{\phi_{\boldsymbol{\alpha}}},\text{DSSM}}
    \bigg(
    V_{\phi_{V}}\left(\mathbf{s}_t\right)-\sum_{\tau=t}^T\gamma^{\tau-t}R_\tau
    \bigg)^2.
    \end{aligned}
\end{equation}
The above losses are then used to update the actor and critic networks via gradient descent.

\subsection{Proposed LocDreamer Learning}
The overall training procedure of LocDreamer consists of two stages: DSSM pre-training and WM imagination-based joint training of DSSM and scheduling policy, as illustrated in \textbf{Algorithm~\ref{alg:pseudocode}}.
\begin{algorithm}[!ht]
    \caption{Proposed LocDreamer Algorithm}\label{alg:pseudocode}
    \SetKwInOut{KwIn}{Input}\SetKwInOut{KwOut}{Output}
    \KwIn{DSSM training epoch $E_\text{dssm}$;
    LocDreamer imagination and training epoch $E_\text{imagine}$;}
    \KwOut{Trained LocDreamer for tracking and anchor scheduling;}
    
    Initialize LocDreamer\;
    \tcc{DSSM Training Using Real Data $\mathcal{R}$}
    \For{$e=1,\,\cdots,\,E_\text{dssm}$}{ \label{alg:line:wm_pretraining:start}
        Initialize $\mathbf{z}_0^\text{posterior}$ and $\mathbf{h}_1$\;
        \For{$t=1,\,\cdots,\,T$}{
            $\mathbf{z}_t^\text{prior} \sim f_{\theta_\mathbf{z}}\left(\mathbf{z}_{t-1}^\text{posterior},\,\mathbf{h}_t\right)$\;
            $\mathbf{z}_{t}^\text{posterior}\sim q_\phi\left(\mathbf{z}_{t-1}^\text{posterior},\,\mathbf{h}_{t},\,\{(d^k_t,\,\mathbf{p}^k)\,|\,k\in \mathcal{R}\}\right)$\;
            $\hat{d}^k_t \sim g_{\theta_{d}}\left(\mathbf{z}^\text{posterior}_t,\,\mathbf{h}_t,\,\mathbf{p}^k\right)$ for $k\in \mathcal{R}$\;
            $\mathbf{h}_{t+1} = r_{\theta_\mathbf{h}}\left(\mathbf{z}^\text{posterior}_{t},\,\mathbf{h}_{t}\right)$\;
            Compute $L_t^\text{dyn}$ from $\mathbf{z}_t^\text{prior},\,\mathbf{z}_t^\text{posterior}$\;
            Compute $L_t^\text{recon}$ from ${d}^k_{t},\,\hat{{d}}^k_t$ for $k\in \mathcal{R}$\;
        }
        Train $\theta_{\mathbf{z}},\,\theta_{\mathbf{h}},\,\theta_{\mathbf{d}},\,\phi$ of DSSM\;\label{alg:line:wm_pretraining:end}
    }        
    \tcc{LocDreamer Imagination and Training for $\mathcal{A}$}
    \For{$e=1,\,\cdots,\,E_\text{imagine}$}{
        Initialize $\mathbf{z}_0^\text{posterior}$ and $\mathbf{h}_1$\;
        \For{$t=1,\,\cdots,\,T$}{
            $\mathbf{z}_t^\text{prior} \sim f_{\theta_\mathbf{z}}\left(\mathbf{z}_{t-1}^\text{posterior},\,\mathbf{h}_t\right)$\;
            ${d}^k_t \sim g_{\theta_{d}}\left(\mathbf{z}^\text{prior}_t,\,\mathbf{h}_t,\,\mathbf{p}^k\right)$ for $k\in\mathcal{A}$\;
            $\boldsymbol{\alpha}_t \sim \pi_{\theta_{\boldsymbol{\alpha}}} \left(\mathbf{z}^\text{prior}_t,\,\mathbf{h}_t\right) \in \mathbb{R}^A$\;
            $\mathbf{z}_t^\text{posterior} \sim q_\phi\left(\mathbf{z}_{t-1}^\text{posterior},\,\mathbf{h}_{t},\, \mathbf{o}\!\left(\boldsymbol{\alpha}_t\right)\right)$\;
            $\hat{d}^k_t \sim g_{\theta_{d}}\left(\mathbf{z}^\text{posterior}_t,\,\mathbf{h}_t,\,\mathbf{p}^k\right)$ for $k\in \mathcal{R}\cup\mathcal{K}_t$\;
            $\mathbf{h}_{t+1} = r_{\theta_\mathbf{h}}\left(\mathbf{z}^\text{posterior}_{t},\,\mathbf{h}_{t}\right)$\;
            Compute $L_t^\text{dyn}$ from $\mathbf{z}_t^\text{prior},\,\mathbf{z}_t^\text{posterior}$\;
            Compute $L_t^\text{recon}$ from ${d}^k_{t}$ and $\hat{d}^k_t$ for $k\in\mathcal{R}\cup\mathcal{K}_t$\;
            Compute AC loss from $L_t^\text{recon}$\;
        }
        Finetune $\theta_{\mathbf{z}},\,\theta_{\mathbf{h}},\,\theta_{\mathbf{d}},\,\phi$ of DSSM and train ${\theta_{\boldsymbol{\alpha}}},\phi_{V}$ of AC for $\mathcal{A}$\;
    }
\end{algorithm}
In the first stage, the DSSM is pre-trained on a source environment with sufficient real measurements from a set of anchors $\mathcal{R}\nsubseteq\mathcal{A}$, assuming all anchors in $\mathcal{R}$ are always providing measurements, i.e., $\boldsymbol{\alpha}_t=\mathbf{1},\,\forall t$.
In each pre-training epoch, the model first initalizes the latent and hidden states, $\mathbf{z}^\text{posterior}_0$ and $\mathbf{h}_1$, and processes the real measurements $\{d_t^r\}_{r\in\mathcal{R}}$ sequentially over $T$ timesteps. At each timestep $t$, the model first predicts the prior latent state $\mathbf{z}_t^\text{prior}$ using~\eqref{eq:target dynamic}, then infers the posterior latent state $\mathbf{z}_t^\text{posterior}$ using~\eqref{eq:encoder}, reconstructs the distance measurements $\{\hat{d}_t^r\}_{r\in\mathcal{R}}$ using~\eqref{eq:decoder}, and finally updates the RNN hidden state $\mathbf{h}_{t+1}$ using~\eqref{eq:rnn sequence}.
At the end of each timestep, the model computes the reconstruction loss $L^\text{recon}_t$ and dynamic loss $L^\text{dyn}_t$ in \eqref{eq:elbo} and accumulates them over $T$ timesteps.
After processing all timesteps in the epoch, the model computes the overall ELBO loss $\mathcal{L}^{\text{ELBO}}$ in~\eqref{eq:elbo} and uses gradient descent to update the DSSM parameters. 
The DSSM pre-training repeats for $E_\text{dssm}$ epochs.

Once trained, the DSSM can model the target dynamics and generate realistic distance measurements for arbitrary anchor deployments.
In detail, in each imagination epoch, the DSSM predicts future dynamics considering the imagined anchors in $\mathcal{A}$ without measurements but only with their positions $\{\mathbf{p}^k\}_{k\in \mathcal{A}}$ known.
At each timestep $t$, the DSSM predicts future dynamics by running learned~\eqref{eq:rnn sequence} and~\eqref{eq:target dynamic} forward, samples imagined measurements from the learned generative model~\eqref{eq:decoder}, and select the anchors using~\eqref{eq:action} accordingly. 
This closed-loop imagination facilitates training entirely inside its dream, allowing LocDreamer to generalize to unseen scenarios without actual interaction with the anchors in $\mathcal{A}$.
After each epoch, the parameters of the DSSM and AC are updated based on their losses. 
The training for the imagined anchors is repeated for $E_\text{imagine}$ epochs.

\section{Experiment Results and Discussions}
\subsection{Experiment Setup}
We evaluate the performance of the proposed framework using the indoor Ultra-wideband (UWB) positioning and position tracking dataset~\cite{Bregar/UWBData/2023/ScientificData}, which uses commercial DecaWave DW1000 UWB modules as measurement nodes operating across frequencies from 3494.4 MHz to 6489.6 MHz with bandwidths of 499.2 MHz or 900 MHz. 
Specifically, we use \textit{Environment 0}, a residential house floor (9.18 m $\times$ 12.06 m) with brick inside and outside walls, and a total of 126410 measurements were collected from 8 fixed anchors at 85 positions.

To mimic a data-scarce deployment scenario, we use real measurements from three anchors to bootstrap the model, and then use the trained model to imagine for unseen set of five new anchors $\mathcal{A}$ with $A=5$.
The tracking model and scheduling policy are trained on these imagined measurements, targeting a fixed minimum number of $K_t=3,\forall t$ anchors. 
Tracking performance is then evaluated using real measurements from the unseen anchors $\mathcal{A}$, testing the framework's generalization to new anchor deployments when trained using imagined data.

All experiments are conducted on a single NVIDIA RTX 3060 Ti GPU and hyperparameters of the model is summarized in Table~\ref{tab:hyperparameter}. 
\begin{table}[!ht]
\renewcommand\arraystretch{0.8}
\centering
\caption{Hyperparameter Setting}
\resizebox{\columnwidth}{!}{
    \begin{tabular}{c|ccc}
    \toprule 
    \textbf{Parameter} & \textbf{Symbol} & \textbf{Value} \\
    \midrule
    \multicolumn{1}{c}{\textbf{Environment}} \\
    \midrule
    Total number of anchors& $A$ & 5 \\
    Scheduled number of anchors & $K_t$ & 3 \\
    \midrule
    \multicolumn{1}{c}{\textbf{DSSM}} \\
    \midrule
    DSSM Training Epoch & $E_\text{dssm}$ & 50 \\
    Batch size & $B$ & 32 \\
    Sequence length & $T$ & 32 \\
    RNN state dimension & $H$ & 50 \\
    RNN layers & - & 2 \\
    Learning rate & - & 1e-3 \\
    Optimizer & - & Adam \\
    Scheduler & - & Cosine annealing \\
    Weight decay & - & 1e-3 \\
    \midrule
    \multicolumn{1}{c}{\textbf{AC}} \\
    \midrule\
    Imagination Epoch & $E_\text{imagine}$ & 300 \\
    Learning rate & - & 1e-3 \\
    Optimizer & - & Adam \\
    Scheduler & - & Cosine annealing \\
    Discount factor & $\gamma$ & 0.99 \\
    \bottomrule
  \end{tabular}
  }
  \label{tab:hyperparameter}
\end{table}
For comparison, we assess the tracking performances of the following methods and strategies:
\begin{enumerate}
    \item EKF - \textit{random scheduling}: 
    Extended Kalman filter (EKF) with ${K}_t=3$ anchors randomly selected at each timestep. 
    A constant velocity transition model~\cite{Wang/DSSM/2025/ICC} is adopted and the standard deviation of acceleration noise for the process noise covariance matrix $\mathbf{Q}\in\mathbb{R}^{4\times4}$ is set to $\sigma_\text{acc}=0.2\text{m/s}^2$.
    The standard deviation of measurement noise for the measurement noise covariance matrix $\mathbf{R}\in\mathbb{R}^{4\times4}$ is set to $\sigma_n=1\text{m}$.
    \item EKF - \textit{all anchors}: 
    EKF with the same configuration but using all anchors $\mathcal{A}$ at each timestep. 
    \item DSSM - \textit{all anchors (real)}:
    Trained with real measurements using all anchors $\mathcal{A}$.
    \item LocDreamer - \textit{random scheduling}: 
    Trained with imagined data and it randomly selects ${K}_t$ anchors instead of using the scheduling policy from learned AC.
    \item LocDreamer - \textit{all anchors (imagined)}:
    Trained with imagined data using all anchors $\mathcal{A}$. 
    \item LocDreamer - \textit{scheduling}:
    Trained with imagined data and scheduling from learned AC. 
\end{enumerate}

\subsection{Tracking Performance}
\begin{figure}[!t]
\centering{\includegraphics[width=0.4\textwidth]{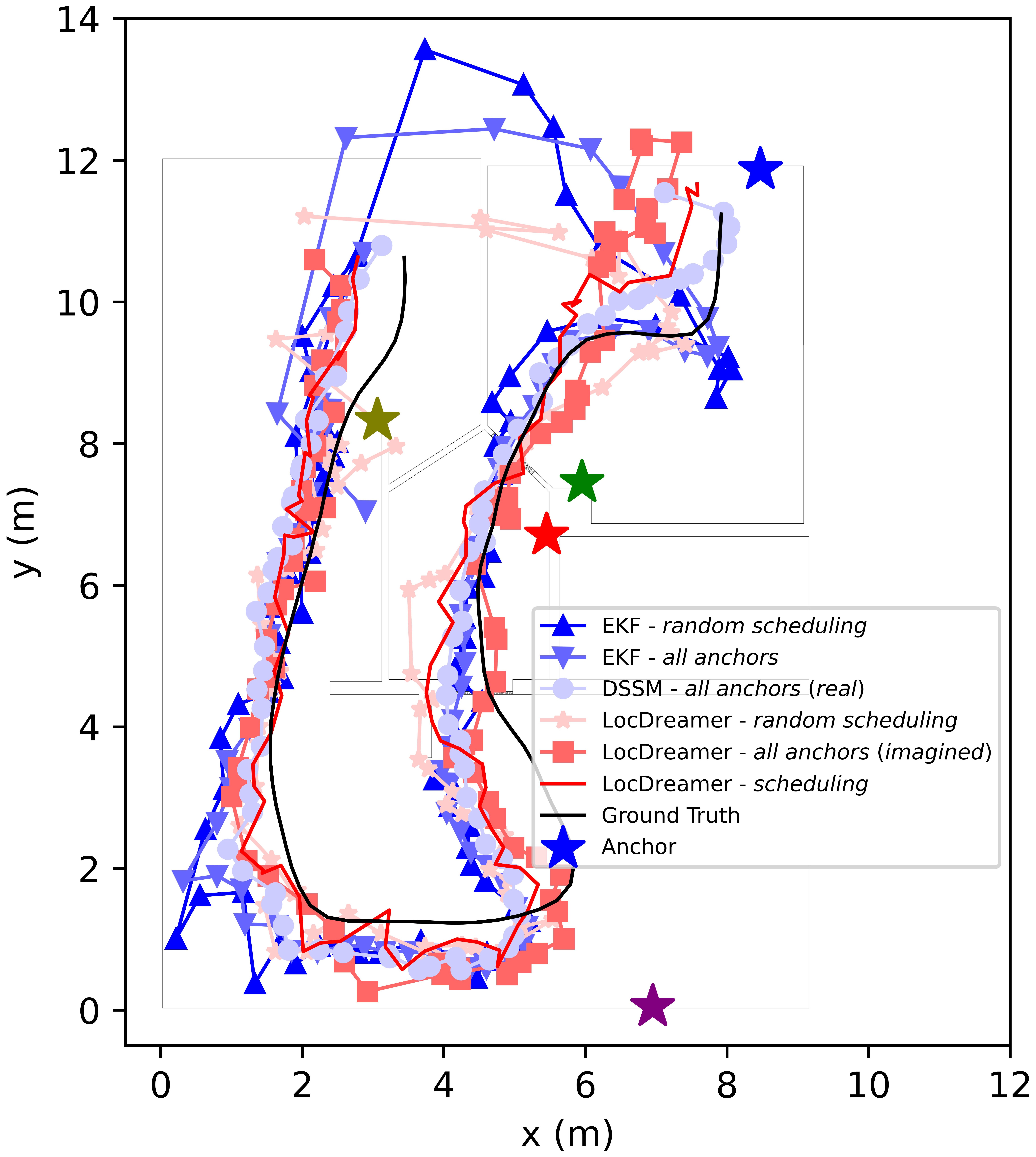}}
	\caption{Estimated trajectories with different methods.}
	\label{fig:trajectories}
    \vspace{-0.2cm}
\end{figure}

We summarize the tracking performances in terms of mean absolute error (MAE), root mean squared error (RMSE), and $50^\text{th}/90^\text{th}$ percentile error for all methods in Table~\ref{tab:tracking performance}. 
\begin{table}[t]
\renewcommand\arraystretch{1.1}
\centering
\caption{Tracking Errors (m) for Different Algorithms}
\resizebox{\columnwidth}{!}{
  \begin{tabular}{c|cccc}
    \toprule 
    Algorithms & MAE & RMSE & $50^{\text{th}}$ & $90^{\text{th}}$\\
      \midrule
      \midrule
      EKF - \textit{random scheduling} & 1.05 & 1.67 & 0.71 & 1.83 \\
      EKF - \textit{all anchors} & 0.92 & 1.50 & 0.63 & 1.41 \\
      DSSM - \textit{all anchors (real)} & 0.57 & 0.64 & 0.54 & 0.96 \\
      LocDreamer - \textit{random scheduling} & 1.07 & 1.43 & 0.82 & 2.05 \\
      LocDreamer - \textit{all anchors (imagined)} & 0.85 & 1.07 & 0.68 & 1.60 \\
      \textbf{LocDreamer - \textit{scheduling}} & \textbf{0.66} & \textbf{0.77} & \textbf{0.60} & \textbf{1.18} \\ 
    \bottomrule
  \end{tabular}
  \label{tab:tracking performance}
  }
\end{table}
EKF and LocDreamer achieve similar tracking performance with a random scheduling strategy, achieving a MAE around 1m. 
By using all anchors instead of random scheduling, the tracking performances increase in all metrics for both EKF and LocDreamer, demonstrating the additional robustness with abundant measurements and validating that LocDreamer can generalize to unseen anchors without requiring real measurements.
The proposed LocDreamer - \textit{scheduling} further improves accuracy while preserving resource efficiency by actively scheduling most informative anchors with policies learned from imagined data, demonstrating that the imagined measurements from LocDreamer are sufficiently realistic to train both the tracking model and the scheduling policy. 
And it approaches the performance of DSSM using all anchors trained with real measurements. 
We also show the estimated trajectories of these methods in Fig.~\ref{fig:trajectories}.
These results demonstrate that LocDreamer can learn latent dynamics and enable imagination-driven learning, leading to efficient and accurate tracking in unseen anchor deployments without real measurements. 

\begin{figure}[!ht]
\centering{\includegraphics[width=0.38\textwidth]{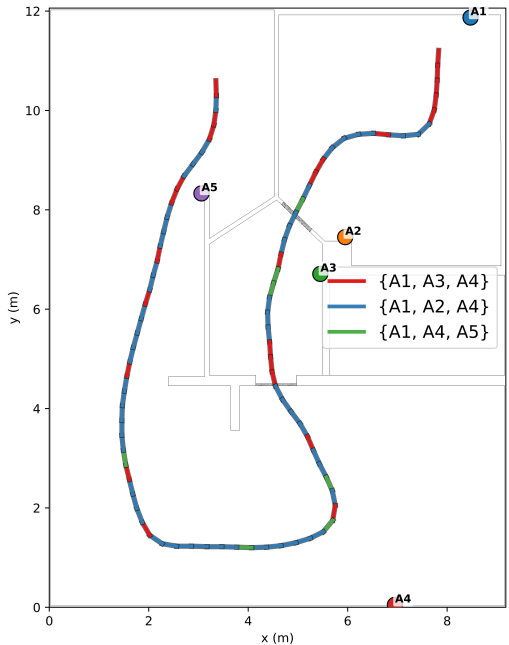}}
	\caption{An anchor scheduling heatmap over spatial locations where different color represents different selected anchor sets $\mathcal{K}_t$.}
	\label{fig:scheduling heatmap}
    \vspace{-0.5cm}
\end{figure}

We also plot an anchor scheduling heatmap across different locations in Fig.~\ref{fig:scheduling heatmap}.
\begin{figure}[ht]
\centering{\includegraphics[width=0.45\textwidth]{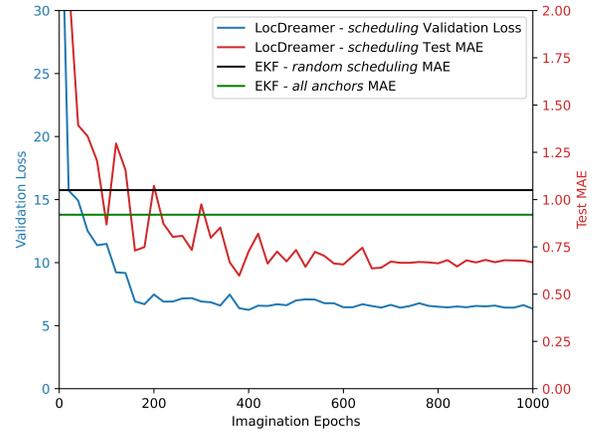}}
	\caption{Validation loss and test MAE of LocDreamer - \textit{scheduling} versus imagination epochs $E_\text{imagine}$.}
	\label{fig:learning curve}
    \vspace{-0.5cm}
\end{figure}
It can be observed that the model consistently prioritizes anchors 1 and 4, which provide favorable geometric diversity and lower geometrical dilution of precision (GDOP)~\cite{Horn/GDOP/2020/Sensors}. 
This confirms that the learned scheduling policy discovers physically meaningful anchor combinations directly from imagination-driven rewards. 
The learning curve of tracking performance against training epochs is shown in Fig.~\ref{fig:learning curve}.
We set $E_\text{imagine}=1000$ to test the LocDreamer's sensitivity to overfitting. 
The proposed LocDreamer - \textit{scheduling} quickly converges at around 300 epochs, outperforming baseline methods. 

\subsection{Limitations and Future Work}
Our current model still assumes real data from a minimal set of 3 anchors to bootstrap training, and the quality of these real data affects the training and imagination performance. 
Reducing the impact or removing this assumption is an important direction for future work. 
Moreover, the current number of scheduled anchors is fixed at $K_t=3$, extending the scheduler to dynamically adapt $K_t$ to channel conditions and desired accuracy-resource tradeoffs is another promising direction. 
The resource efficiency/usage can be modeled in a more sophisticated way for quantitative evaluation. 
Finally, further validation on large-scale and more complex environments is a promising next step. 
In these scenarios, exhaustive search becomes computationally intractable, and the proposed imagination-driven framework could demonstrate clear advantages in scalability, data efficiency and cross-environment generalization. 

\section{Conclusion}
This paper introduces LocDreamer, a WM-based learning framework for joint indoor tracking and anchor scheduling.
By learning a DSSM that captures target dynamics and environment behavior, the proposed approach can imagine realistic measurements for unseen anchor deployments.
These imagined measurements enable efficient training of both the tracking model and a RL-based scheduling policy, without requiring additional data from site surveys. 
Experiments on a public UWB dataset demonstrate that the proposed framework achieves superior accuracy using only a few anchors for bootstrapping, outperforms random scheduling and approaches the performance of models trained with full real measurements.
Future work will explore adaptive anchor selection with variable anchor budgets and self-bootstrapping that requires no real data initialization to further enhance robustness and scalability toward AI-native 6G networks. 

\section{Acknowledgments}
This work was supported by Australian Research Council (ARC) under Grants DP210103410 and DP220101634, in part by SUTD Kickstarter Initiative (SKI 2021\_06\_08), and in part by the National Research Foundation, Singapore and Infocomm Media Development Authority under its Communications and Connectivity Bridging Funding Initiative.

\bibliography{aaai2026}


\end{document}